\documentclass[12pt,a4]{article}
\pdfoutput=1
\usepackage{graphicx}
\usepackage{amsmath}
\usepackage{subfigure}
\usepackage{gensymb}
\usepackage{color}
\usepackage{xcolor}
\usepackage{soul}
\usepackage{braket}
\usepackage{hyperref}
\usepackage[left=1.5in, right=1.5in, top=1.25in, bottom=1.25in]{geometry}
\date{}

\begin{document}

\title { \bf Independent measurement of Muon neutrino and anti-neutrino oscillations at the INO-ICAL Experiment}

\author{ Zubair Ahmad Dar$^{\ast}$, Daljeet Kaur$^{\dagger}$, Sanjeev Kumar$^{\ddagger}$, Md. Naimuddin$^{\ddagger}$ \\
$^{\ast}$Aligarh Muslim University, Aligarh\\
$^{\dagger}$S.G.T.B. Khalsa College, University of Delhi\\
$^{\ddagger}$Department of Physics and Astrophysics, University of Delhi}

\maketitle

\begin{abstract}

The magnetised Iron Calorimeter detector at the India-based Neutrino Observatory (INO) has a unique feature to identify the neutrinos and antineutrinos on an event by event basis.
This feature can be harnessed to detect the differences between the oscillation parameters of neutrinos and antineutrinos independently.
In this paper, we analysed Charged Current $\nu_{\mu}$ and $\overline{\nu}_{\mu}$ events under the influence of earth matter effect using three
neutrino flavor oscillation framework. If the atmospheric mass-squared differences and mixing parameters for neutrinos are different from antineutrinos, we present 
the prospects for the experimental observation of these differences in atmospheric $\nu$ and $\overline \nu_{\mu}$ oscillations at INO. We estimate the detector sensitivity
to confirm a non-zero  difference in the mass-squared splittings ($|\Delta m^{2}_{32}|-|\Delta\overline{m^{2}}_{32}|$) for neutrinos and antineutrinos. 
\end{abstract}

\newpage 
\section{Introduction}
 The phenomenon of neutrino oscillation is well established by many experiments involving solar\cite{SNO1}, atmospheric \cite{SK1,SK2}, 
accelerator \cite{K2K}, and reactor neutrinos\cite{KAM1}. It exhibits that neutrino flavor eigenstates 
($\nu_{e}, \nu_{\mu}, \nu_{\tau}$) are indeed quantum superpositions of mass eigenstates ($\nu_{1}, \nu_{2}, \nu_{3}$) with definite 
masses ($m_{1}, m_{2}, m_{3}$)
represented mathematically as 
\begin{eqnarray}
 \ket{\nu_{i}}  &=\sum_{\alpha}U_{\alpha i}\ket{\nu_{\alpha}}
\end{eqnarray}
where,
$\ket{\nu_{i}}$ represents a neutrino with a definite mass $m_{i}$ (i=1,2,3), $\ket{\nu_{\alpha}}$ represents a neutrino with a definite flavor, and $U_{\alpha i}$ is the famous Pontecorvo Maki
Nakagawa-Sakata (PMNS) lepton mixing matrix \cite{bruno, pmns1}. The oscillation probability depends on three mixing angles, $\theta_{12}, \theta_{13}, \theta_{23}$; two mass differences, 
$\Delta m^{2}_{21}$=$ m^{2}_{2}-m^{2}_{1}$, and $\Delta m^{2}_{31}$=$m^{2}_{3}-m^{2}_{1}$ , and a CP phase $\delta_{CP}$.

 Although a remarkable progress has been made by several neutrino experiments to measure these oscillation parameters with reasonable accuracy \cite{global,Reno,DB,T2K}, still there are several physics concerns that perhaps lie beyond paradigm of the three-massive-neutrinos scheme. 
The particles and their antiparticles are assumed to have equal masses and their different couplings are closely related as a consequence of the CPT-theorem.
Therefore, parameters governing neutrino and antineutrino oscillation probabilities are considered to be identical. But, there is a possibility that neutrino and antineutrino may behave differently \cite{Bar1,Bar2,Bar3, Bar4,raj,raj1,poonam}. The survival probability for muon neutrinos at a particular energy $E_{\nu}$ and propagation length $L$ is given by
\begin{equation}
P(\nu_{\mu} \rightarrow \nu_{\mu}) \simeq 1-4\cos^{2}\theta_{13}\sin^{2}\theta_{23} \times [1-\cos^{2}\theta_{13}\sin^{2}\theta_{23}]\sin^{2}(\frac{1.267 |\Delta m^{2}_{32}|L}{E_{\nu}}).
\end{equation}
Similarly, the survival probability for muon antineutrino i.e., P($\bar\nu_{\mu} \rightarrow \bar\nu_{\mu}$) can be written by replacing the 
neutrino parameters by the corresponding antineutrino parameters which are denoted mathematically by placing a bar on neutrino parameters.

Comparing the oscillation parameters of neutrinos and antineutrinos could, therefore, be a particular test of CPT-conservation or any difference between them may indicate a sign of new physics.
Some experiments such as MINOS \cite{minos11,minos12,minos_proc} and Super-Kamiokande (SK) \cite{SK_paper} have performed some analyses with their experimental data  assuming non-identical parameters for neutrinos and anti-neutrinos and found that neutrino and antineutrinos oscillation parameters are in agreement. Also, the magnetized Iron Calorimeter (ICAL) 
detector of the India-based Neutrino Observatory (INO)\cite{INO3} can easily distinguish an atmospheric $\nu_{\mu}$ and $\overline\nu_{\mu}$ events on an event by event basis with its excellent charge identification capability due presence of a strong magnetic field. A detail of the ICAL detector at the INO is given in Sec. \ref{secino}. This paper presents the future ICAL
sensitivity for the measurement of muon neutrino and antineutrino oscillation parameters assuming that neutrinos and antineutrinos have 
different atmospheric mass-squared splittings and mixing angles assuming Normal mass Hierarchy(NH) is true. We study the prospects of the scenario when both the differences ($|\Delta m^{2}_{32}|-|\Delta\overline{m^{2}}_{32}|$) and 
($\sin^{2}\theta_{23}$-$\sin^{2}\overline{\theta}_{23}$) are non-zero. Earlier INO study as in Ref \cite{prd} shows the ICAL detector sensitivity to measure the difference  ($|\Delta m^{2}_{32}|-|\Delta\overline{m^{2}}_{32}|$) when only mass square splittings of neutrinos and anti-neutrinos are different with the assumption that $\nu_{\mu}$ and $\bar\nu_{\mu}$  mixing angles are identical i.e.($\sin^{2}\theta_{23}$-$\sin^{2}\overline{\theta}_{23}$=0). In this paper, with the realistic detector resolutions and efficiencies of the ICAL, we vary all the four atmospheric oscillation parameters 
($|\Delta m^{2}_{32}|, |\Delta\overline{m^{2}}_{32}|, \sin^{2}\theta_{23}$, $\sin^{2}\overline{\theta}_{23}$) simultaneously for neutrinos and antineutrinos to get a four dimensional fit. Using the results of this four parameters fit analysis, we show the ICAL detector potential to observe the difference between the neutrino and antineutrino mass-squared splittings ($|\Delta \overline{m^{2}}_{32}|-|\Delta m^{2}_{32}|$) and its sensitivity for ruling out the identical oscillation parameter hypothesis.

\section{The INO-ICAL Experiment}
\label{secino}
The India-based Neutrino Observatory (INO) is an atmospheric neutrino experimental facility that will be located in Southern India. An Iron-Calorimeter 
(ICAL) will be the prime detector at INO to address the current issues in neutrino physics. The aim of the ICAL detector is to observe 
the neutrino and anti-neutrino oscillations separately using $\nu_{\mu}$ ($\overline\nu_{\mu}$) disappearance channel with good precision in GeV 
energy range. The detector is expected to be magnetized to about 1.5 T, allowing differentiation of the events induced by muon neutrinos and 
muon antineutrinos. 
 Through this sensitivity, one can probe the difference in matter effects in the propagation of neutrinos and antineutrinos traversing through
 the Earth. This, in turn, will allow for a sensitivity to the neutrino mass hierarchy, which is one of the primary goals of the ICAL experiment.
 The ICAL detector will consist of three modules, each module will have a dimension of 16 m $\times$ 16 m $\times$ 14.5 m, comprising a total 
 weight of about 50 kton. Each module will be a stack of 150 layers, where 5.6 cm thick iron plates are interleaved with Resistive Plate Chambers (RPCs) of dimension  2m $\times$ 2m having gas thickness of 3 mm. A total of 30000 RPCs \cite{DJ_nim} are going to be used as active detector element for the INO-ICAL detector. 

A typical Charged Current interaction of $\nu_{\mu}$ (or $\overline \nu_{\mu}$) with the iron target produces a charged muon and single 
or multiple hadrons.  Muon deposit their energy in iron forming a clear track-like pattern while hadrons form a shower or cluster like pattern. 
The good tracking ability and energy resolution of ICAL for muons make it very well suited for the study of neutrino oscillation physics and
in addition, its sensitivity to multi-GeV hadrons provides a significant improvement in its physics potential \cite{DJ_epjc,hdphy}.

\section{Methodology}

The magnetized ICAL detector enables separation of neutrino and antineutrino interactions for atmospheric events, allowing an independent measurement of the neutrino and antineutrino
oscillation parameters. Here, we analyze the reach of the Iron Calorimeter for neutrino and antineutrino oscillations separately using a three flavor analysis including the Earth 
matter effects. We use a large number of unoscillated NUANCE\cite{nuance} generated neutrino events, with an exposure of 50 kt $\times$ 1000 years of the ICAL detector, and then 
finally normalize to 500 kt-yr. We use HONDA\cite{honda} atmospheric neutrino fluxes for event generation. Each CC neutrino event is characterized by its energy and zenith angle.

\begin{figure}[htbp]
  \centering

   \subfigure[]{
  \includegraphics[width=0.45\textwidth,height=5.99cm]{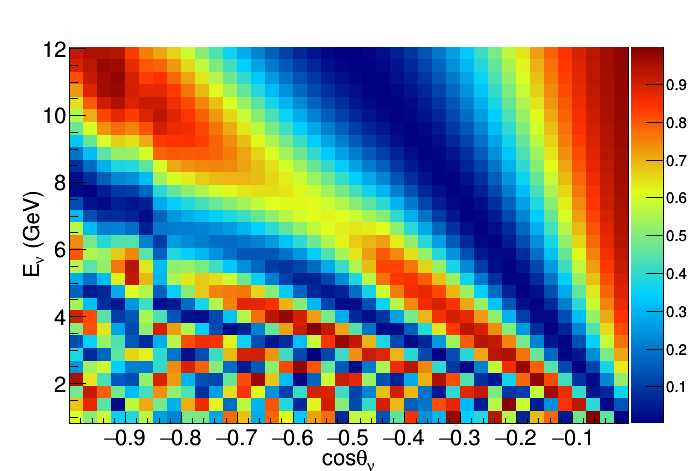}
   \label{nu_osc}
   }
 \subfigure[]{
  \includegraphics[width=0.45\textwidth,height=6cm]{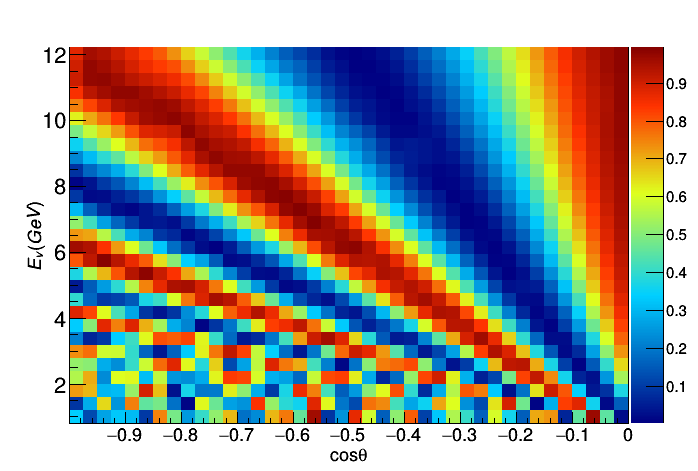}
   \label{anu_osc}
   }

\caption{\label{oscillogram} Oscillograms for muon neutrino (Left) and anti-neutrino (Right) survival probabilities on E-$\cos\theta$ palne including the Earth matter effects using $|\Delta m^{2}_{23}|(or|\Delta \bar m^{2}_{23}|)$=2.4 $\times$ $10^{-3} eV^{2}$ and $\sin^2\theta_{23} (or\sin^2\bar\theta_{23}) $=0.5}
\end{figure}

Table~\ref{osc_tb1} shows the oscillation parameters which are kept fixed throughout the analyses presented in this paper. The solar 
oscillation parameters ( $\Delta m^{2}_{21}$ and $ \sin^{2}\theta_{12}$) are kept fixed, as they do not show any significant impact on the results.  
As $\theta_{13}$ is now known quite precisely, it has been fixed as well. Since, the ICAL is insensitive to the variation of $\delta_{CP}$ phase
\cite{cpsense}, hence it is also fixed at $0^{\degree}$. Oscillation effects have been introduced via a Monte-Carlo reweighting algorithm as described in
earlier works \cite{DJ_epjc,trk,moonmoon}. Figure \ref{oscillogram} shows oscillograms for $\nu_{\mu}$ and $\bar\nu_{\mu}$ survival probabilities assuming Normal Hierarchy is true. It is clear from the figure that due to the presence of the Earth matter effect,  $\nu_{\mu}$ and $\bar\nu_{\mu}$ oscillations are different. The charge sensitive ICAL detector can easily distinguish the $\nu_{\mu}$ and $\bar\nu_{\mu}$ oscillations and hence can easily measure their oscillation parameters separately with good precision.   

\begin{table}[htbp]
\begin{center}
\begin {tabular}{c  c  c}
\hline
\hline
Oscillation parameters & True values & Marginalization range  \\
\hline 
\hline

 $ \sin^2\theta_{13}$ &  0.0234 & Fixed \\              
$\sin^2\theta_{12} (or\sin^2\bar\theta_{12}) $ & 0.313 & Fixed \\
 $\Delta m^{2}_{12}(or\Delta \bar m^{2}_{12})$ (eV$^{2}$) & 7.6 $\times$ $10^{-5}$ & Fixed\\
 $\delta_{CP}$ & 0.0 & Fixed \\
\hline

\end {tabular}
\caption{\label{osc_tb1} True values of the neutrino/antineutrino oscillation parameters used in the analysis.} 
\end{center}

\end{table}

Each oscillated neutrino or antineutrino event is divided as a function of twenty muon energy bins ($E_{\mu}$), twenty muon zenith angle ($\cos\theta_{\mu}$) and five hadron 
energy bins ($E_{hadron}$) of optimized bin width as mentioned in Ref.\cite{prd}. These binned data are then folded with detector efficiencies and
resolution functions as provided by the INO collaboration \cite{mureso,hreso} for the reconstruction of neutrino and antineutrino events 
separately.

Though the INO-ICAL have  very good charge identification efficiency, it is still possible that some muon events (say $\mu^{-}$) are misidentified as of opposite charge particles (say $\mu^{+}$) and vice versa. This misidentification of events has been taken care using following procedure as mentioned in references\cite{trk, INO3}.  
Due to the mis-identification, the total number of events, reconstructed as $\mu^{-}$ will increase by
\begin{equation}
\label{eq:effeq}
N^{\mu^{-}}= N^{\mu^{-}}_{RC}+(N^{\mu^{+}}_{R}- N^{\mu^{+}}_{RC}),
\end{equation}
where $N^{\mu^{-}}$ is the number of total reconstructed $\mu^{-}$ events.
$N^{\mu^{-}}_{RC}$ is the number of $\mu^{-}$ events reconstructed and correctly identified in charge and $N^{\mu^{+}}_{RC}$ is the same for $\mu^{+}$ events with their respective reconstruction and charge identification efficiencies folded in; whereas $N^{\mu^{+}}_{R}$ is the number of reconstructed $\mu^{+}$ events.
Hence, $N_{R}-N_{RC}$ gives the fraction of reconstructed events that have their charge wrongly identified.  All the quantities given in Eq.\ref{eq:effeq} are function of $E_{\mu}$ and $\cos\theta_{\mu}$ and are determined bin wise. Total rightly identified reconstructed $\mu^{+}$ events can be obtained using similar expression with charge reversal.

We use a ``pulled'' $\chi^{2}$ \cite{maltoni,sys3,sys4} method based on Poisson probability distribution to compare the expected and observed data
with inclusion of systematic errors (a 20\% error on atmospheric neutrino flux normalization, a 10\% error on neutrino cross-section, an overall 5\% systematic error, 
a 5\% uncertainty due to zenith angle dependence of the fluxes, and an energy-dependent tilt error), as considered in earlier ICAL analyses \cite{INO3, trk, moonmoon, sys1}. All systematic uncertainities are correlated and the first two listed systematic errors should cover the difference between neutrinos and anti-neutrinos.

 The systematic uncertainties and the theoretical errors are parameterized in terms of a set of variables $\zeta$, called pulls.  
 Due to the fine binning, we use the poissonian log likelihood ratio given as,
 
 \begin{equation}
\label{eq:chieq}
 \chi^2(\nu_{\mu}) = min\sum_{i,j,k}\left(2 (N^{T^{\prime}}_{ijk}(\nu_{\mu})-N^{E}_{i,j,k}(\nu_{\mu}))+2N^{E}_{i,j,k}(\nu_{\mu})
(\ln\frac{N^{T}_{i,j,k}(\nu_{\mu})}{N^{T^{\prime}}_{i,j,k}(\nu_{\mu})})\right)+ \sum_{n}\zeta^{2}_{n}, 
 \end{equation}
 where 
\begin{equation}
\label{eq:evteq}
  N^{T^{\prime}}_{ijk}(\nu_{\mu}) = N^{T}_{i,j,k}(\nu_{\mu})\left(1 + \sum_{n}\pi^{n}_{ijk}\zeta_{n}\right). 
 \end{equation}
 Here, $N^{E}_{ijk}$ are the observed number of reconstructed events, generated using true values of the oscillation parameters  in $i^{th}$ muon energy bin, $j^{th}$ muon direction bin 
 and $k^{th}$ hadron 
 energy bin, $N^{T}_{ijk}$ are the number of theoretically predicted events generated by varying oscillation parameters, $N^{T^{\prime}}_{ijk}$ show modified events spectrum due to
 different systematic uncertainties, $\pi^{n}_{ijk}$ are the systematic shift in the events of the respective bins due to $n^{th}$ systematic error. The univariate pull variable $\zeta_{n}$, 
  corresponds to the $\pi^{n}_{ijk}$ uncertainty. 
An expression similar to Eq.(~\ref{eq:chieq}) can be obtained for $\chi^{2}(\overline{\nu}_{\mu})$ using reconstructed $\mu^{+}$ event samples.

The functions $\chi^2(\nu_{\mu})$ and $\chi^2(\overline{\nu}_{\mu})$ are calculated separately for the independent measurement of neutrino and antineutrino oscillation parameters. All the systematic uncertainities are correlated and applied to neutrino and anti-neutrino events separately. Each $\chi^2$ is fitted with 20 muon energy bins, 20 muon angle bins and 5 hadron energy bins via $20\times20\times5 =2000$ binning scheme for 
neutrino as well as for antineutrinos. The two $\chi^{2}$ can be added to get the 
combined
$\chi^2(\nu_{\mu}+\overline{\nu}_{\mu})$ as
 \begin{equation}
\label{eq:chiino}
  \chi^2(\nu_{\mu}+\overline{\nu}_{\mu}) =\chi^{2}(\nu_{\mu}) + \chi^{2}(\overline{\nu}_{\mu}).
 \end{equation}

To estimate the ICAL sensitivity for the measurement of oscillation parameters, in the full parameter space, we vary all atmospheric oscillation
parameters ($|\Delta m^{2}_{32}|, |\Delta\overline{m^{2}}_{32}|, \sin^{2}\theta_{23}$ and $\sin^{2}\overline{\theta}_{23}$) 
in their allowed ranges as mentioned in Table~\ref{osc_tb2}. The Charged Current (CC) $\nu_{\mu}$ and $\overline{\nu}_{\mu}$ events spectrum are separately binned into direction and
energy bins. The $\chi^{2}$ function is minimized with respect to these four parameters along with the five nuisance parameters to take into account the systematic uncertainties
for different energy and direction bins.

After performing feasibility study, we perform our analysis in two steps:\\
(1) Observed values of all four oscillation parameters ($|\Delta m^{2}_{32}|$, $\sin^{2} \theta_{23}$, $|\Delta \overline {m^{2}}_{32}|$, $\sin^{2}\overline\theta_{23}$) are varied within in an experimentally allowed range as given in Table~\ref{osc_tb2} keeping their true values fixed and non-identical.\\
(2) The true values of all the four oscillation parameters ($|\Delta m^{2}_{32}|$, $\sin^{2} \theta_{23}$, $|\Delta \overline {m^{2}}_{32}|$, $\sin^{2}\overline \theta_{23}$) are varied in a wide range and a $\chi^{2}$ has been calculated to find out sensitivity for non-identical mass-squared splittings and mixing angles of $\nu_{\mu}$ 
and $\overline\nu_{\mu}$.

\begin{table}[htbp]
\begin{center}
\begin {tabular}{c c}
\hline
\hline
oscillation parameters &  Range  \\
\hline 
\hline
$|\Delta m^{2}_{32}|$ (eV$^{2}$) & (2.0-3.0) $\times$ $10^{-3}$\\
$|\Delta \overline{m^{2}}_{32}|$ (eV$^{2}$) & (2.0-3.0) $\times$ $10^{-3}$  \\
$ \sin^2\theta_{23}$ & 0.3-0.7 \\
$ \sin^2\overline {\theta}_{23}$ &   0.3-0.7 \\
\hline 
\end {tabular}
\caption{\label{osc_tb2} The neutrino and antineutrino oscillation parameters and their experimentally allowed range used in the analysis.} 
\end{center}

\end{table}

 \subsection{Feasibility study}
\begin{figure}[htbp]
 \centering

  \includegraphics[width=0.7\textwidth,height=6cm]{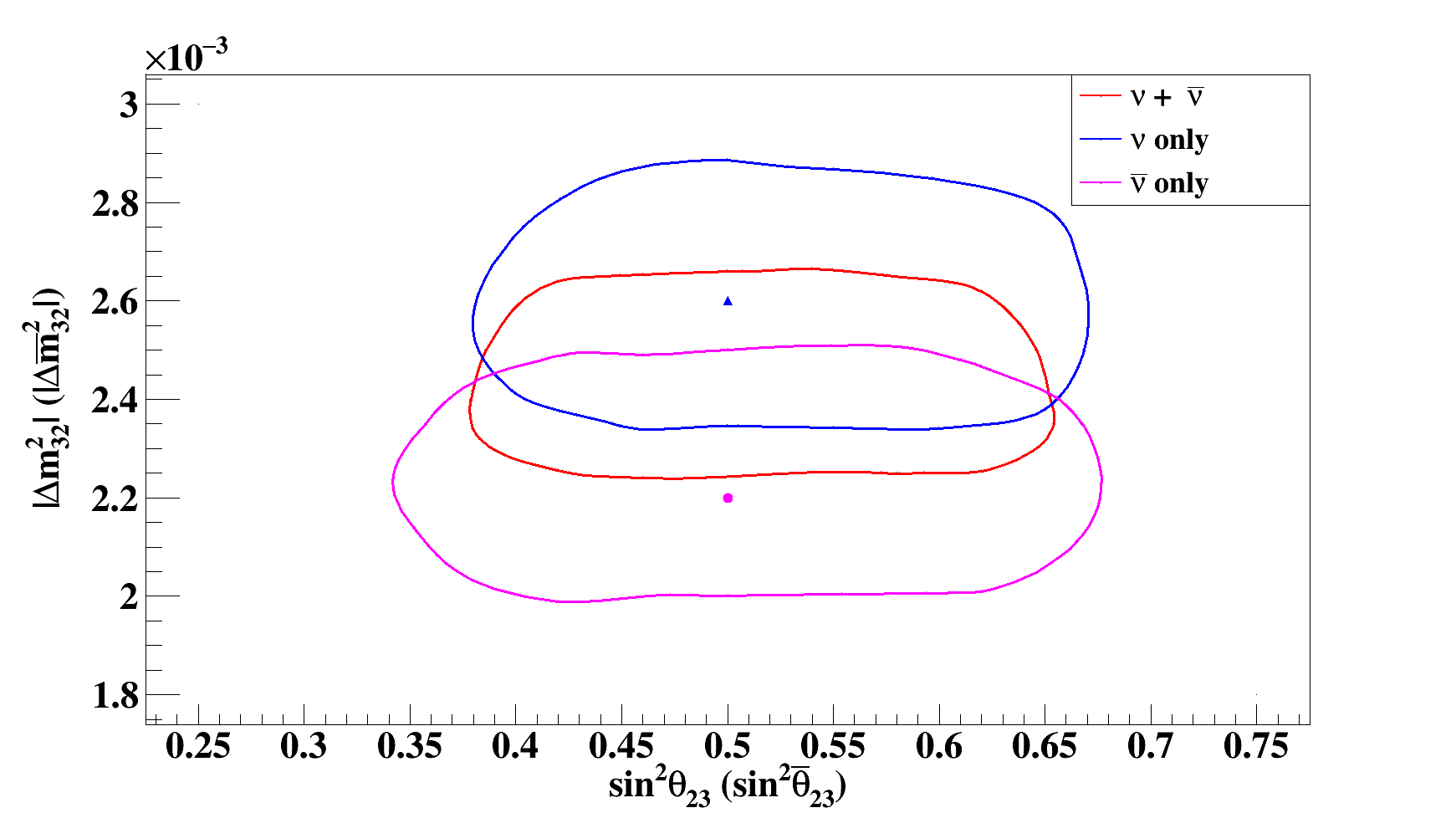}

\caption{\label{figA} 99$\%$ C.L. contours obtained from $\chi^{2}(\nu)$, $\chi^{2}(\overline\nu)$ separately and with combined $\chi^{2}$using non-identical true values of mass
splittings for neutrinos ($|\Delta m^{2}_{32}|=2.6 \times 10^{-3} eV^{2}$) and for antineutrino ($|\Delta \overline{m^{2}}_{32}|=2.2 \times 10^{-3} eV^{2}$).}
\end{figure}

  Lets consider a scenario where neutrino and antineutrino have different oscillation parameters. We generate the INO-ICAL events for the oscillation parameters 
  as shown in Table~\ref{osc_tb1} with the assumption that neutrinos and antineutrinos have different mass-squared splittings. We
  use ($|\Delta m^{2}_{32}|=2.6 \times 10^{-3}(eV^{2})$ and $|\Delta \overline {m^{2}}_{32}| = 2.2 \times 10^{-3} (eV^{2})$. These $\nu$ and $\overline\nu$ events are then binned
  into $\cos\theta_{\mu}$, and E$_\mu$ and E$_{hadron}$ bins separately. Further, $\chi^{2}(\nu)$ and $\chi^{2}(\overline\nu)$ have been calculated separately and
  we show the  99$\%$ Confidence Level (C.L.) contours for ($|\Delta m^{2}_{32}|$, $\sin^{2} \theta_{32}$) and  
  ($|\Delta \overline{m^{2}}_{32}|$, $\sin^{2}\overline\theta_{32}$) in 
  Figure ~\ref{figA}. The contours in blue and magenta show the sensitivity of INO for $\nu_{\mu}$ and $\overline \nu_{\mu}$ respectively for the scenario where they have
  different atmospheric mass-squared splittings.

  However, if the combined $\chi^{2}$ is calculated as mentioned in Eq.~\ref{eq:chiino}, with the observation that neutrino and antineutrino have identical oscillation
  parameters although their true values are different then this sensitivity is shown with a red contour in Figure ~\ref{figA}. It is clear that such a combined $\chi^{2}$ 
  analysis will give the best fit value that is more precise compared to that obtained from $\chi^{2}_{\nu}$ and $\chi^{2}_{\overline\nu}$ separate analyses.  
  We calculate the precision as $\frac{P_{max}-P_{min}}{P_{max}+P_{min}}$, where $P_{max}$ and $P_{min}$ are the maximum and minimum limits at the given CL of the corresponding oscillation parameters on the given axis. We find that the precision of the combined analysis is improved as expected [Table~\ref{pre_tb}]. But, it is highly likely that in this case the unrealistic sensitivity may be
  obtained when the difference between $\nu_{\mu}$ and $\overline \nu_{\mu}$ is ignored. 
  As it is clear from Figure ~\ref{figA} that the best fit obtained from combined $\chi^{2}$ analyses is roughly an average value between the true values of $\Delta m^{2}$ and $\Delta \overline m^{2}$ which is less 2$\sigma $ away from the given true values of neutrino and antineutrino mass squared splittings.

  %a compromised value between two true values from  $\chi^{2}_{\nu}$ and 
  %$\chi^{2}_{\overline\nu}$ analyses as 
  
  Thus, in order to achieve the accurate sensitivity without ignoring the 
  difference between oscillation parameters or to test the hypothesis that neutrinos and antineutrinos share the identical parameters, we should  allow for the 
  possibility of different true values of $\nu_{\mu}$ and $\overline \nu_{\mu}$ parameters in nature. For this, we need to vary the true as well as observed values
  of all four parameters i.e. ($|\Delta m^{2}_{32}|$, $\sin^{2} \theta_{23}$, $|\Delta \overline {m^{2}}_{32}|$, $\sin^{2}\overline\theta_{23}$) in the analyses.

  \begin{table}
\begin{center}
\begin {tabular}{c c c}
\hline
\hline
Analysis & $ \sin^2\theta_{23}$ &$|\Delta m^{2}_{32}|$ (eV$^{2}$)  \\
\hline 
\hline
Neutrino events &  28.84$\%$ &10.66$\%$ \\
Anti-neutrino events &  32.66$\%$ & 14.51$\%$\\
Combined($\nu_{\mu}+\overline{\nu}_{\mu}$)&  26.57$\%$ & 8.57$\%$\\
 \hline
\end {tabular}
\caption{\label{pre_tb}Precision values at the 99$\%$ C.L. considering different oscillation parameters for neutrino, antineutrino and with the 
combined ($\nu_{\mu} +\overline{\nu}_{\mu}$) events [as shown in Figure ~\ref{figA}] assuming $\sin^2\theta_{23}=\sin^2\overline{\theta}_{23}$ and $|\Delta m^{2}_{32}|$=$|\Delta\overline{m^{2}}_{32}|$ (eV$^{2}$) for $\chi^{2}$ calculations } 
\end{center}
\end{table}

   \subsection{Measurement with the Non-identical, fixed true values}
  \subsubsection{ Four-parameter fit and extraction of two-parameter fit}
  This study has been performed to extract the sensitivity of the ICAL detector on a four parameter space assuming non identical parameters 
  for $\nu_{\mu}$ and $\overline\nu_{\mu}$. Here, $\chi^{2}$ have been 
  calculated as a function of four atmospheric oscillation parameters ($|\Delta m^{2}_{32}|$, $ \sin^2\theta_{23}$,  $|\Delta \overline{m}^{2}_{32}|$, 
  $ \sin^2\overline\theta_{23}$) while all other oscillation parameters are kept fixed at their central values.

 \begin{figure}[htbp]
 \centering

  \includegraphics[width=0.7\textwidth,height=6cm]{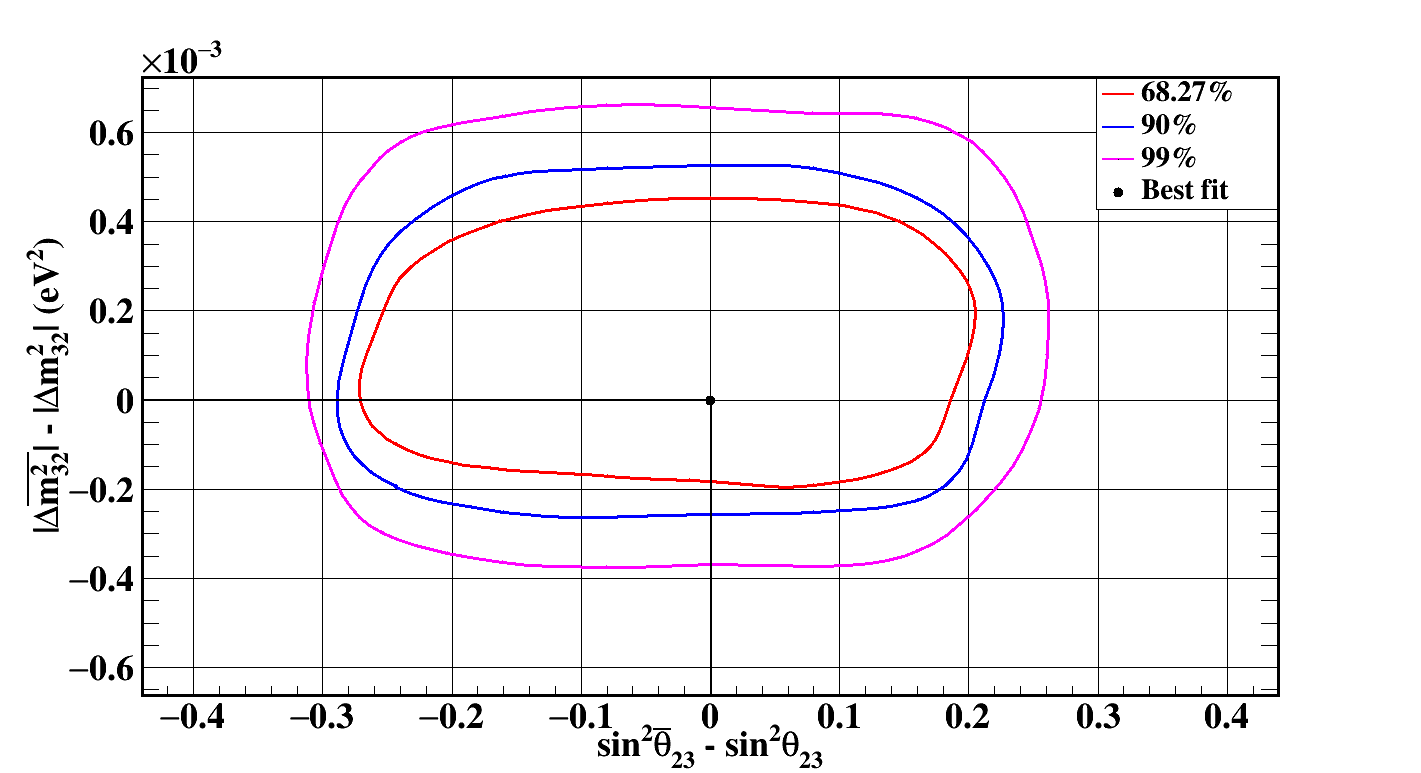}

\caption{\label{figB} The ICAL sensitivity on ($\delta_{m}=|\Delta\overline{m^{2}}_{32}|-|\Delta m^{2}_{32}|)$) and ($\delta_{\theta}=\sin^{2}\overline{\theta}_{23}-\sin^{2}\theta_{23}$) plane at $68\%$, $90\%$ and $99\%$ confidence levels. Origin is the point where, neutrino and antineutrino parameters are identical.}
\end{figure}

\begin{figure}[htbp]
  \centering
  
  \includegraphics[width=0.7\textwidth,height=6.4cm]{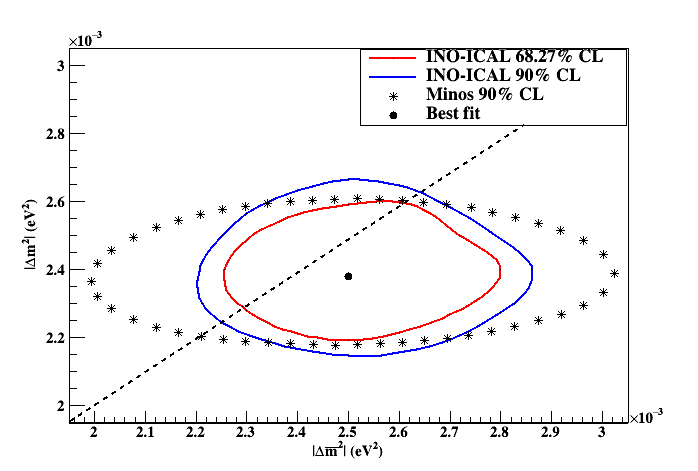}
   
   \caption{\label{minoscomp} The 68$\%$ and 90$\%$ confidence level contours on the $|\Delta m^{2}_{32}|$ and $|\Delta \overline{m^{2}}_{32}|$ parameter space showing the sensitivity of the ICAL experiment using atmospheric data only and Minos experiment using combined beamline and atmospheric data as given in Ref.\cite{minos_proc}. Dashed line shows $|\Delta m^{2}_{32}|=|\Delta \overline{m^{2}}_{32}|$ }
\end{figure}

We start with the assumption that neutrino and antineutrinos have different mass-squared splittings but identical mixing angles as
$|\Delta m^{2}_{32}|$=$2.38 \times 10^{-3} (eV^{2})$, $|\Delta \overline{m}^{2}_{32}|$=$2.5 \times 10^{-3} (eV^{2})$ such that the difference
$(|\Delta\overline{m^{2}}_{32}|-|\Delta m^{2}_{32}| =0.12)$, and $ \sin^2\theta_{23}$= $ \sin^2\overline\theta_{23}$=0.5 such that
$(\sin^{2}\overline{\theta}_{23}-\sin^{2}\theta_{23}=0)$.
A fake dataset is generated at the given fixed true values of oscillation parameters ($|\Delta m^{2}_{32}|$, $ \sin^2\theta_{23}$,  $|\Delta \overline{m}^{2}_{32}|$, $ \sin^2\overline\theta_{23}$). A four dimensional grid search ( $10 \times 5 \times 10 \times 5$ ) is performed for the predicted dataset. $\chi^2$ is calculated between the fake dataset and predicted dataset for each set of oscillation parameters.
%It is to be noted that the true values of all the oscillation parameters are fixed while observed four parameters
%($|\Delta m^{2}_{32}|$, $ \sin^2\theta_{23}$,  $|\Delta \overline{m}^{2}_{32}|$, $ \sin^2\overline\theta_{23}$) are varied simultaneously in a grid of a $10 \times 5$
%matrix of discrete values in the neutrino plane and a $10 \times 5$ matrix in
%the antineutrino plane within their given  marginalization range as 
%shown in Table~\ref{osc_tb2}.

The $\chi^{2}$ for neutrino and antineutrino has been calculated separately, and a combined $\chi^{2}$ 
sensitivity is considered for the estimation of the differences in mass-squared splittings  
($\delta_{m}=|\Delta\overline{m^{2}}_{32}|-|\Delta m^{2}_{32}|)$ and mixing angles ($\delta_{\theta}=\sin^{2}\overline{\theta}_{23}-\sin^{2}\theta_{23}$) of 
neutrinos and antineutrinos.
Figure~\ref{figB} plots the differences between the oscillation parameters on 
($\delta_{m}=|\Delta\overline{m^{2}}_{32}|-|\Delta m^{2}_{32}|$) and  ($\delta_{\theta}=\sin^{2}\overline{\theta}_{23}-\sin^{2}\theta_{23}$) plane
at $68\%$, $90\%$ and $99\%$ confidence levels.
In general, there will be several points from the four dimensional $\chi2$ surface but a minimum $\chi^2$ has been chosen among those points to take the final single value in that bin.

A set of two parameters profile can also be extracted from the four parameters $\chi^{2}$ data set by minimizing with respect to pairs of remaining oscillation parameters.
Figure ~\ref{minoscomp} shows the ICAL sensitivity for atmospheric mass-squared splitting on $|\Delta m^{2}_{32}|$ and $|\Delta\overline{m^{2}}_{32}|$ parameter space
by minimizing over $\sin^{2}\overline{\theta}_{23}$ and $\sin^{2}\theta_{23}$ at different confidence intervals. It is clear from the figure that the ICAL can measure  
$|\Delta m^{2}_{32}|$ and $|\Delta\overline{m^{2}}_{32}|$ with a precision of about 10.41$\%$ and 12.87$\%$ at 90$\%$ Confidence Levels, respectively. The diagonal dashed line in  Figure ~\ref{minoscomp} indicates the case of identical mass splittings and 
mixing angles for neutrinos and antineutrinos, respectively.
The neutrino mass-squared splittings on the $|\Delta m^{2}_{32}|$ and $|\Delta \overline{m^{2}}_{32}|$ parameter space at different confidence intervals obtained from 
MINOS detector using both beamline and atmospheric data has been shown in Figure 4 of Ref.~\cite{minos_proc},  having similar fixed true values as mentioned in Table~\ref{osc_tb}. 
 Figure ~\ref{minoscomp} shows that using similar oscillation parameters, the ICAL 
sensitivity for neutrinos is almost comparable to that of MINOS as shown in Ref.~\cite{minos_proc} while qualitatively, the ICAL is more sensitive than MINOS for the antineutrinos.

   \subsubsection{ICAL sensitivity in $\delta_{m}$ and $\delta_{\theta}$ plane with non-identical true parameters }
\begin{figure}[htbp]
 \centering
 \subfigure[]{
  \includegraphics[width=0.45\textwidth,height=6cm]{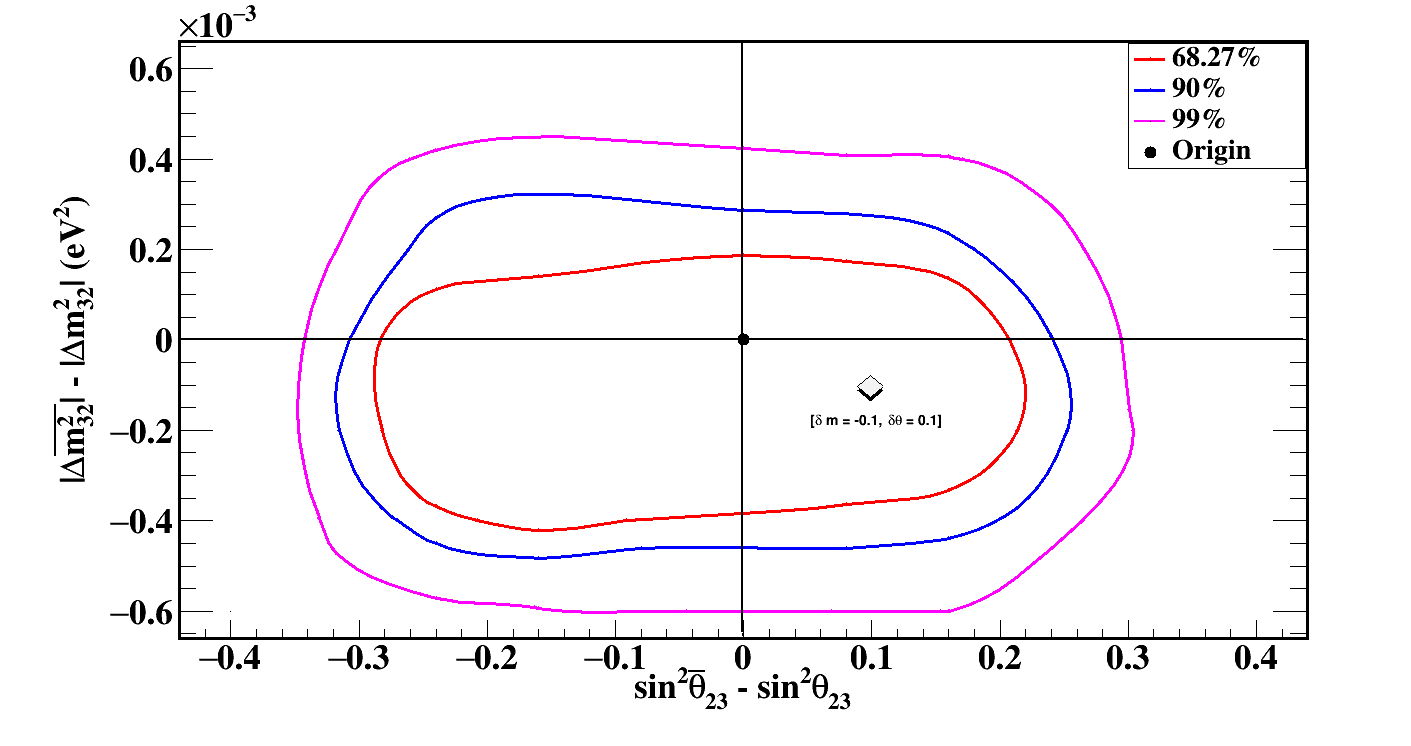}
   \label{fig:a}
   }
 \subfigure[]{
  \includegraphics[width=0.45\textwidth,height=6cm]{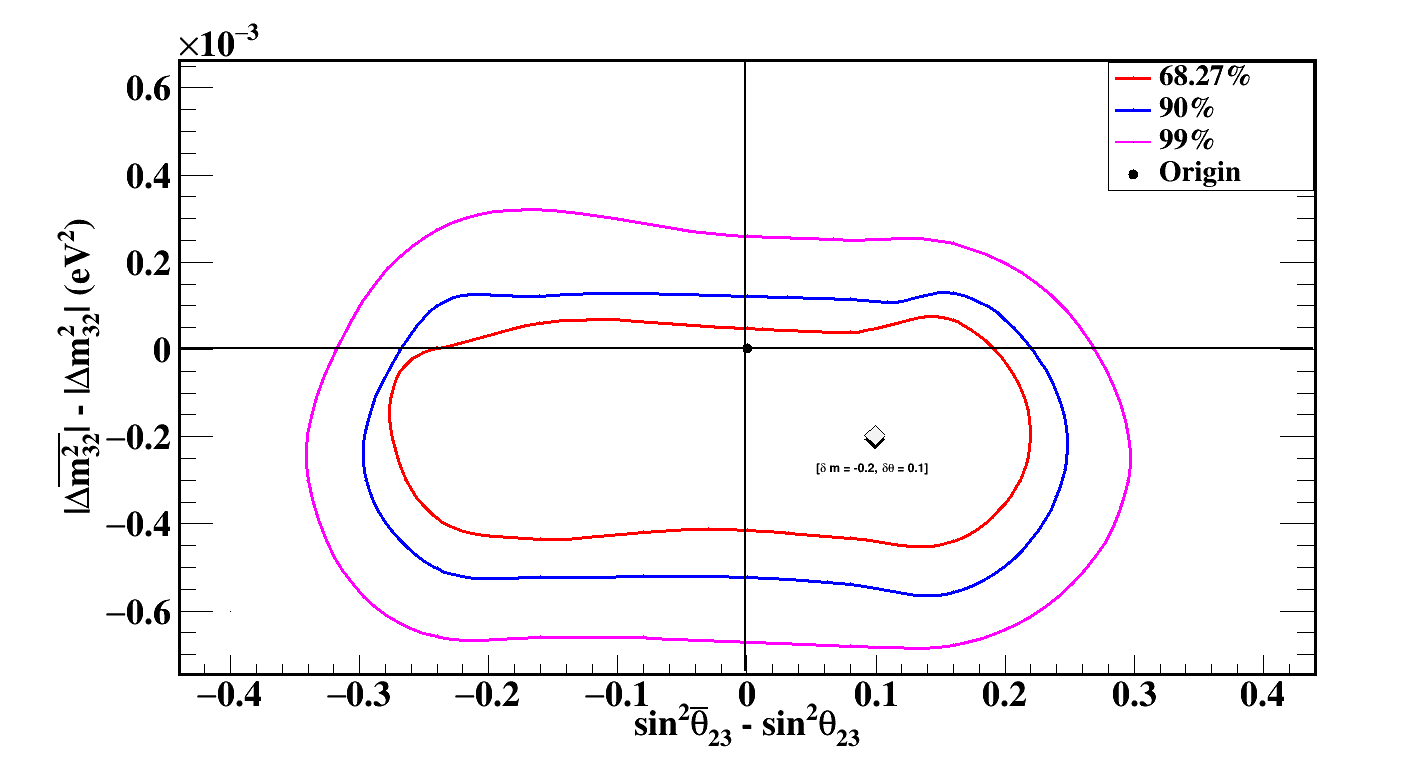}
   \label{fig:b}
   }
\subfigure[]{
  \includegraphics[width=0.45\textwidth,height=6cm]{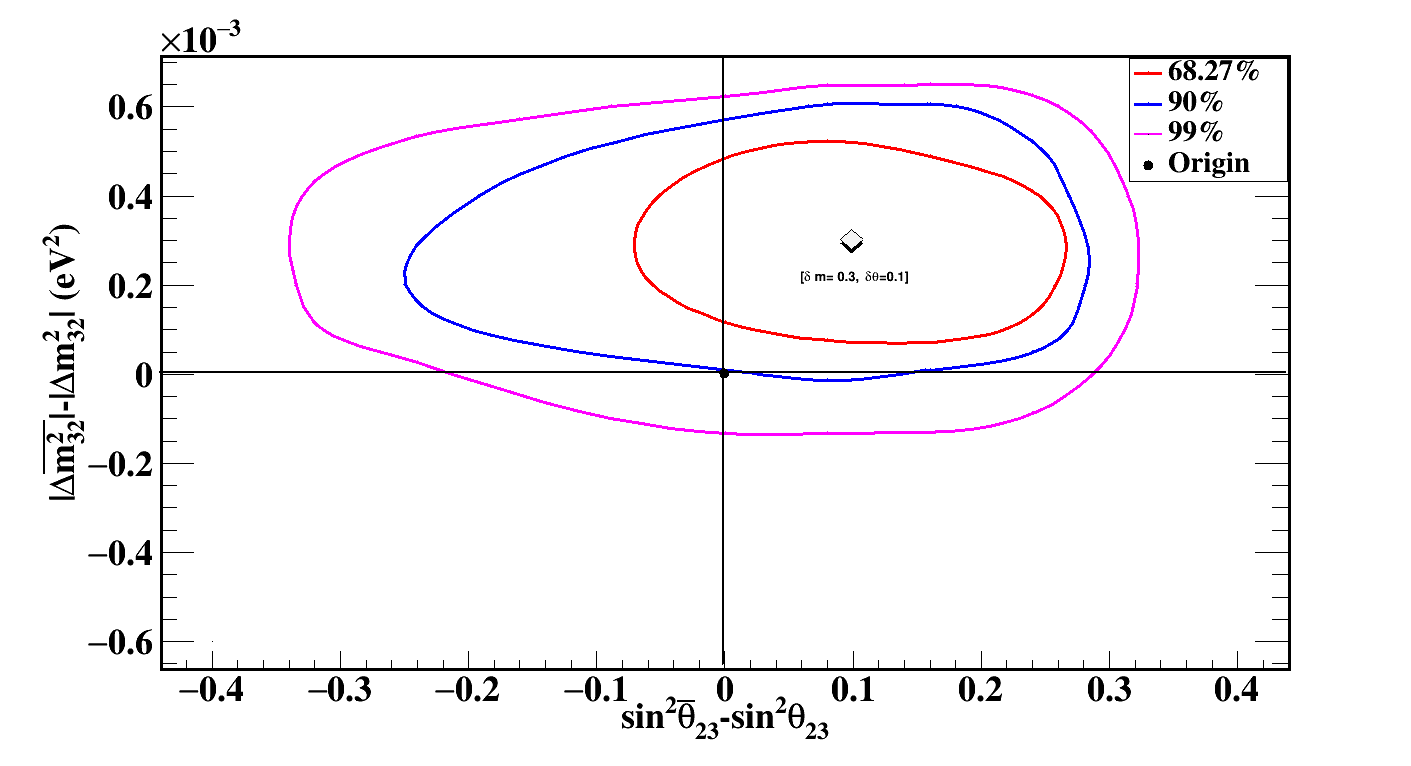}
   \label{fig:c}
   }
\subfigure[]{
  \includegraphics[width=0.45\textwidth,height=6cm]{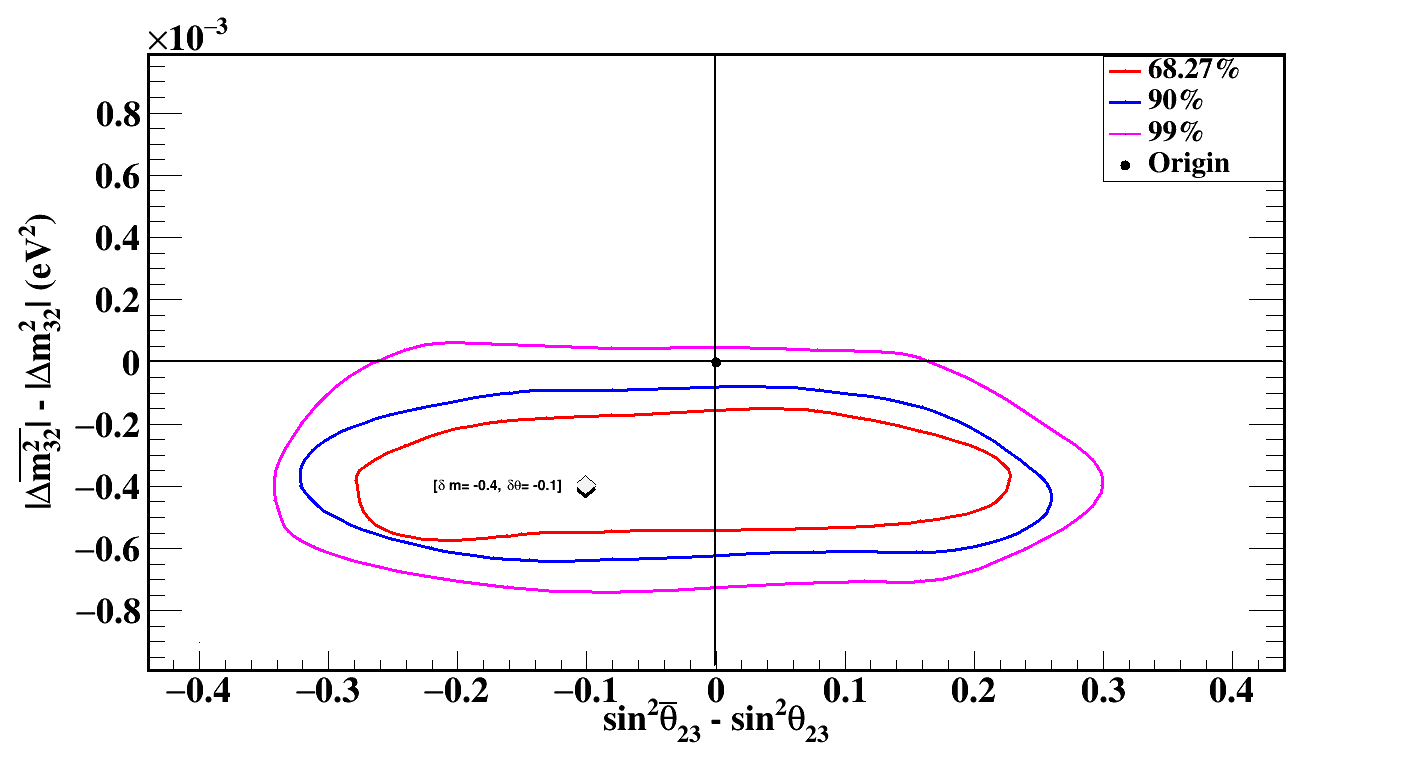}
   \label{fig:d}
   }
\caption{\label{sample}Contour plots at 68$\%$, 90$\%$ and 99$\%$ C.L. for different true values of $|\Delta m^{2}_{32}|$ and $ |\Delta\overline{m^{2}}_{32}|$ as mentioned 
in Table ~\ref{osc_tb}. 
Here, X-axis corresponds to the differences of $ \sin^2\theta_{23}$ and $\sin^{2}\overline{\theta}_{23}$ and Y-axis corresponds to differences in  $|\Delta m^{2}_{32}|$ and 
$|\Delta\overline{m^{2}}_{32}|$ values. In these plots diamond shows the best fit value of the observed parameters.}
\end{figure}

 Further, we performed the similar four fit $\chi^{2}$ study for different sets of fixed, but non-identical true values of atmospheric oscillation parameters 
 to check the ICAL sensitivity to rule out the hypothesis that neutrinos and antineutrinos have identical oscillation parameters.
  Figure ~\ref{sample} shows the sample sensitivity plots for different combinations of oscillation parameters as shown in Table~\ref{osc_tb} as a function of 
  ($\delta_{m}$) and  ($\delta_{\theta}$) at different Confidence Levels (C.L.). In these plots, the origin point shows the null hypothesis where neutrino and antineutrino
  parameters could be identical or in other words ($\delta_{m}=\delta_{\theta}=0$). It can be seen from these figures that as $\delta_{m}$ and $\delta_{\theta}$ move 
  away from the origin point either in the positive or negative direction (as shown in Table~\ref{osc_tb}), the ICAL sensitivity to the null hypothesis varies significantly. 
  For example, Figure ~\ref{fig:a} and \ref{fig:b} having [$\delta_{m},\delta_{\theta}$] as [$-0.1\times 10^{-3}$, 0.1] and  [$-0.2\times 10^{-3}$, 0.1] shows that using 
  the corresponding mass-squared splitting and mixing angles for $\nu$ and $\overline\nu$, the ICAL can rule out the null hypothesis only at less than $1\sigma$ ($68\%$)level. 
  Similarly, Figure ~\ref{fig:c} and \ref{fig:d} having [$\delta_{m},\delta_{\theta}$] as [$0.3\times 10^{-3}$, 0.1] and  [$-0.4\times 10^{-3}$,-0.1] shows the same 
  at $2\sigma$ ($90\%$) and more than  $2\sigma$ level.
 Hence, to estimate the real significance of the ICAL detector for ruling out the null hypothesis or to reveal any mismatch in the 
 $\nu$ and $\overline\nu$ parameters, it is pertinent to vary the true values of all four oscillation fit parameters rather than fixing them at any certain value, as is
  done in the next section.

\begin{table}[htbp]
\begin{center}
\begin {tabular}{c|c|c|c|c|c|c}
\hline

\hline
Set No. &$|\Delta m^{2}_{32}|(eV^{2})$&$|\Delta \overline{m^{2}}_{32}|(eV^{2})$&$\sin^2\theta_{23}$&$ \sin^2\overline{\theta}_{23}$&$\delta_{m}(eV^{2})$&$ \delta_{\theta}$\\ \hline
Set-1&$2.6\times10^{-3}$&$2.5\times10^{-3}$&0.5&0.6&$-0.1\times 10^{-3}$&0.1 \\
Set-2&$2.6\times10^{-3}$&$2.4\times10^{-3}$&0.5&0.6&$-0.2\times 10^{-3}$&0.1 \\
Set-3&$2.2\times10^{-3}$&$2.5\times10^{-3}$&0.4&0.5&$0.3\times 10^{-3}$&0.1 \\
Set-4&$2.4\times10^{-3}$&$2.0\times10^{-3}$&0.5&0.4&$-0.4\times 10^{-3}$&-0.1 \\
 \hline
\end {tabular}
\caption{\label{osc_tb} True values of the neutrino and antineutrino mass-squared splittings, mixing angles and their differences used in the analysis} 
\end{center}

\end{table}

\section{ICAL potential for non-identical mass-squared splittings}
\label{diff_sens}

In this section, the true as well as observed values of atmospheric oscillation parameters (i.e.$|\Delta m^{2}_{32}|$, $|\overline{\Delta m^{2}}_{32}|,\sin^{2}\theta_{23}$
and $\sin^{2}\overline{\theta}_{23}$) have been allowed to vary independently as given in Table~\ref{osc_tb2}. The ICAL sensitivity to validate a non-zero value of the differences in $\nu$ and $\overline\nu$ mass-squared splittings ($\delta_{m} \neq 0 $), true values of oscillation parameters are set to be non-identical. These true values are also  varied simultaneously in a grid of $6\times 5$ for neutrino plane and $6\times 5$ for anti-neutrino plane. Further, we assume 
the identical parameters for neutrinos and antineutrinos ($\delta_{m}$ =0) and ($\delta_{\theta}$ =0) as our null hypothesis.
To test this null hypothesis, we estimate the $\chi^{2}(\nu+\overline{\nu})$ only for observed ($|\Delta m^{2}_{32}|$=$|\overline{\Delta m^{2}}_{32}|$) and 
($\sin^{2}\theta_{23}=\sin^{2}\overline{\theta}_{23}$) values. The $\chi^{2}$ is calculated for each set of true values of $|\Delta m^{2}_{32}|$, $\sin^{2}\theta_{23}$ ,
$|\overline{\Delta m^{2}}_{32}|$, and $\sin^{2}\overline{\theta}_{23}$. 

\begin{figure}[htbp]
 \centering
  \includegraphics[width=0.7\textwidth,height=7cm]{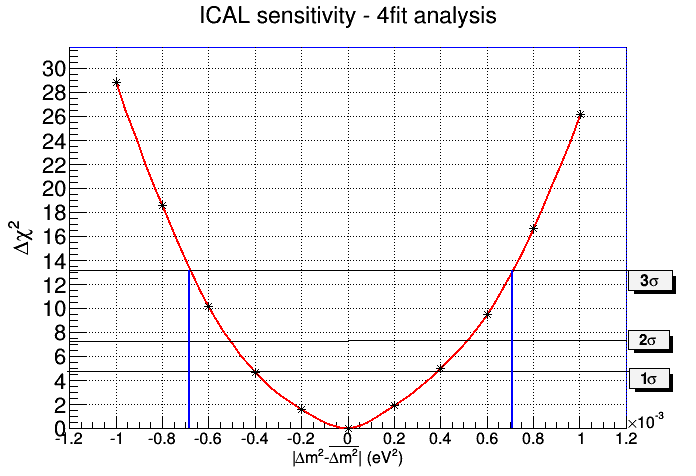}
 
\caption{\label{trueplot2}The INO-ICAL sensitivity for the difference between true values of mass-squared splittings of neutrinos and anti-neutrinos $(|\Delta m^{2}_{32}|-|\Delta\overline{m^{2}}_{32}|)_{True}(eV^{2})$ at 1$\sigma$ ($68\%$), 2$\sigma$ ($90\%$) and 3$\sigma$ (99$\%$) obtained with minimising over true values of other two oscillation parameters i.e. $\sin^{2}\theta_{23}, \sin^{2}\overline{\theta}_{23}$.  
confidence
levels using four oscillation parameter fit technique.}
\end{figure}

 A minimum $\chi^2$ has been binned as a function of difference in the true values of $[\delta_{m}]_{True}$ keeping marginalization over 
 $[\sin^{2}\theta_{23} $ and $\sin^{2}\overline{\theta}_{23}]_{True}$.
 %Similarly,  a minimum $\chi^2$ is calculated as a function of difference in the true   values of $[\delta_{\theta}]_{True}$ keeping marginalization over $[|\Delta m^{2}_{32}|$ and $ |\overline{\Delta m^{2}}_{32}|]_{True}$.
 %This results in several  $\chi^2$ points corresponding to a common set of differences of mass-squared and mixing angles. For each set of difference  $[\delta_{m}]_{True}$  and  
 %$[\delta_{\theta}]_{True}$, we calculate $\Delta\chi^{2}=\chi^{2}-\chi^{2}_{min}$ and plot it as the functions of set of differences. Figure ~\ref{trueplot2} 
 %represents  the sensitivity of the ICAL for $[|\Delta m^{2}_{32}|-|\Delta\overline{m^{2}}_{32}|]_{True}$ with minimisation over true values of other two oscillation parameters ($\sin^{2}\theta_{23},\sin^{2}\overline{\theta}_{23}$). This represents the INO-ICAL potential for ruling out the null hypothesis $|\Delta m^{2}_{32}|=|\Delta\overline{m^{2}}_{32}|$.
This results in several  $\chi^2$ points corresponding to a common set of differences of mass-squared splittings. For each set of difference  $[\delta_{m}]_{True}$  and  
 , we calculate $\Delta\chi^{2}=\chi^{2}-\chi^{2}_{min}$ and plot it as the functions of set of differences. Figure ~\ref{trueplot2} 
 represents  the sensitivity of the ICAL for $[|\Delta m^{2}_{32}|-|\Delta\overline{m^{2}}_{32}|]_{True}$ with minimisation over true values of other two oscillation parameters ($\sin^{2}\theta_{23},\sin^{2}\overline{\theta}_{23}$). This represents the INO-ICAL potential for ruling out the null hypothesis $|\Delta m^{2}_{32}|=|\Delta\overline{m^{2}}_{32}|$. 

\section{Results and Conclusions}

The INO-ICAL potential for the distinct measurements of neutrino and antineutrino oscillation parameters for ten years of exposure
have been investigated. It is shown that to get the accurate sensitivity of the ICAL detector and to test the  hypothesis that neutrinos and antineutrinos share the identical parameters, the difference between oscillation parameters can not be ignored. Therefore, we allow the possibility of different true values of $\nu_{\mu}$ and $\overline \nu_{\mu}$ parameters ($|\Delta m^{2}_{32}|$, $\sin^{2} \theta_{23}$, $|\Delta \overline {m^{2}}_{32}|$, $\sin^{2}\overline\theta_{23}$) in nature. With four parameters fitting analyses, using fixed but different true values of four oscillation parameters, we have shown the ICAL sensitivity for the measurement of the differences $|\Delta m^{2}_{32}|-|\Delta\overline{m^{2}}_{32}|$. Further, 
the extraction of two parameter plots from four parameters fit provides sensitivity for individual oscillation parameters. It has been found that ICAL can measure $|\Delta m^{2}_{32}|$ and $|\Delta\overline{m^{2}}_{32}|$ with a precision of about $\sim 10 \%$ and $\sim 13 \%$ at 90$\%$ Confidence Levels, respectively. Qualitatively, we found that the ICAL is slighltly better sensitive for the anti-neutrinos mass-squared splittings compared to the MINOS as presented in Ref.~\cite{minos_proc}, by using the atmospheric events only while for the neutrinos mass-squared splitting, its sensitivity is almost similar to that of MINOS.

Further, we investigate the scenario where the neutrino and antineutrino oscillation parameters have different true values. We measure the ICAL sensitivity for ruling out the null hypothesis $(|\Delta m^{2}_{32}|=|\Delta\overline{m^{2}}_{32}|)$ by estimating the difference between the true values of mass-squared differences of neutrinos and antineutrinos i.e. $(|\Delta m^{2}_{32}|-|\Delta\overline{m^{2}}_{32}|)$. We find that ICAL can rule out the null hypothesis of $|\Delta m^{2}_{32}|=|\Delta\overline{m^{2}}_{32}|$ at more than 3$\sigma$ ($99\%$)level if the difference of true values of $|\Delta m^{2}_{32}|-|\Delta\overline{m^{2}}_{32}|\geq +0.7\times10^{-3}eV^{2}$ or $|\Delta m^{2}_{32}|-|\Delta\overline{m^{2}}_{32}|\leq -0.7\times10^{-3}eV^{2}$.

\section{Acknowledgement}
 We thank Department of Science and Technology (DST), India and University of Delhi R$\&$D grants for providing the financial support for this research. One of the author also thanks DST-SERB for providing the financial
 support under project No. EMR/2016/002285 to carry out this work.

\end{document}